\documentclass{jpsj2}
\title{Perturbation Theory on the Superconductivity of Heavy
Fermion Superconductor ${\rm UPd_{2}Al_{3}}$}

\author{Yunori \textsc{Nisikawa}\thanks{nisikawa@sci.osaka-cu.ac.jp} and
Kosaku \textsc{Yamada}}

\inst{Department of Physics, Kyoto University, Kyoto 606-8502}

\abst{We investigate the superconducting mechanism and the transition temperature of heavy
fermion superconductor ${\rm UPd_{2}Al_{3}}$ on
the basis of a single band two-dimensional Hubbard model on triangular lattice, which
represents the most heavy band of ${\rm UPd_{2}Al_{3}}$.
Both normal and anomalous self-energies are calculated up to
third order with respect to the Coulomb repulsion $U$ between itinerant
electrons. The superconducting
transition temperature is obtained by solving the
${\rm\acute{E}}$liashberg's equation.
Reasonable transition temperatures are obtained for moderately large $U$. It is found that
the momentum and frequency dependence of
spin fluctuations given by RPA-like terms gives rise to the d-wave
pairing state, while
the vertex correction terms are important for obtaining reasonable transition temperatures.
These results seem to show that the superconductivity in
 ${\rm UPd_{2}Al_{3}}$ can be explained by the perturbation
theory with respect to $U$.}

\kword{heavy fermion systems, electron correlation, ${\rm UPd_{2}Al_{3}}$,
superconducting transition temperature,
two-dimensional Hubbard model on triangular lattice, perturbation
approach, vertex correction
}

\begin{document}
\maketitle
\section{Introduction}
\subsection{Introduction to ${\rm UPd_{2}Al_{3}}$}
${\rm UPd_{2}Al_{3}}$ crystallizes in the hexagonal ${\rm
PrNi_{2}Al_{3}}$-type structure and it was found to be a superconductor
with $T_{\rm c}=$2K. The superconducting state coexists  with antiferromagnetic order below the
N${\rm\acute{e}}$el temperature $T_{\rm N}$=14.5K~\cite{rf:tyoudenAFexp}. The
magnetic moments are aligned ferromagnetically in the hexagonal basal
plane and coupled antiferromagnetically between planes with wave vector ${\bf
k}=[0,0,0.5]$~\cite{rf:magstruc}.

For $T \le T_{\rm N}$, the electrons in ${\rm UPd_{2}Al_{3}}$ behave as if they were separated 
into two rather independent subsystems of 5f character~\cite{rf:pmeasure,rf:muon}.
One subsystem is a localized system which is responsible for
antiferromagnetic order. We call this system ``localized subsystem''.
The second subsystem is the itinerant one characterized
as a heavy electron system and responsible for the superconductivity.
We call the latter ``itinerant subsystem''.
For $T_{\rm c}<  T \ll T_{\rm N}$ in the normal state, the electronic specific-heat $C$
can be approximated by $C\simeq \gamma T$ where $\gamma=145$mJ/${\rm
K}^{2}$~\cite{rf:tyoudenAFexp}.
For $T_{\rm c}<T<$4K in the normal state, the electrical resistivity curve
follows the dependence $\rho=\rho_{0}+AT^{2}$ where $\rho_{0}=2.3{\rm
\mu\Omega cm}$ and $A\simeq10^{-1}{\rm \mu\Omega cm/K^{2}}$~\cite{rf:resitherm}.
The behavior observed in the electronic specific-heat and the electrical 
resistivity indicates that the itinerant electrons in ${\rm
UPd_{2}Al_{3}}$ compose a Fermi liquid.

The calculated Fermi surface in the antiferromagnetic
state has four principal sheets of predominately
$f-$electron character~\cite{rf:band1,rf:band2,rf:band3ref,rf:yamaga}. These are corroborated by
de Haas-van Alphen measurements~\cite{rf:band3ref}. 
All bands are doubly degenerated, because the
antiferromagnetic structure is conserved under a time-reversal and 
inversion-symmetry operator with the non-primitive translation vector
${\bf v}=(0,0,0.5)$. The Fermi sheets are not of hexagonal
symmetry, reflecting the antiferromagnetic structure.
Roughly speaking, two Fermi sheets have two-dimensional topology,
and the remaining two Fermi sheets have the dispersion along the
crystallographic $c-$axis,
although there exist some discrepancies between the results of ref.\citen{rf:band2} and ref.\citen{rf:band3ref} .
In ref.\citen{rf:band2}, two Fermi sheets which have two-dimensional topology are called
, ``party hat'' and ''cylinder'', and the remaining two
are called ``cigar'' and ''eggs''.
According to the band calculation and de Haas-van Alphen measurements~\cite{rf:band3ref},
the band which has the largest effective mass is ``party hat''.

As for the superconducting property, the specific heat and the nuclear
spin-lattice relaxation rate $T_{1}$ show the $T^{3}$- dependence without 
the coherence peak just below $T_{\rm c}$~\cite{rf:NMR1,rf:NMR,rf:NMRd}. 
The Knight shifts decrease below $T_{\rm c}$~\cite{rf:NMR1}.
The thermal conductivity $\kappa$ dose not show any anomaly at 
$T_{\rm c}$ and shows the $T^{2}$ behavior below 1K~\cite{rf:resitherm}.
These facts suggest that the superconductivity is in principle of a d-wave
 type characterized by vanishing gaps on the lines on the Fermi surface,
although there exist some discrepancies between the results of the specific heat
and NMR.
Inelastic neutron scattering
~\cite{rf:neutron1,rf:neutron2,rf:neutron2d,rf:neutron3,rf:neutron3d,rf:neutron3dd}
and tunneling experiments~\cite{rf:tonnelexp1,rf:tonnelexp2}
below the superconducting transition temperature have been
performed on this material. 
The results of these experiments indicate that superconductivity and
magnetic excitations of antiferromagnetically ordered moments in
${\rm UPd_{2}Al_{3}}$ couple each other.
\subsection{Motivation}
Itinerant electrons in heavy fermion systems  compose
a Fermi liquid whose effective mass is enhanced by $10^{2}\sim 10^{3}$ times
as large as that of free electron.
The electron correlation is essential in giving the large electron mass
 in heavy electron systems.
From this point of view, we can expect that superconductivities in heavy fermion
systems are derived also from the electron correlation
through the momentum and frequency dependence of the effective
interaction between electrons.
In fact, in other strongly correlated electron systems such as 
high-$T_{\rm c}$ cuprates~\cite{rf:HoTOPT},
organic superconductors~\cite{rf:JuTOPT} and  spin-triplet
superconductor ${\rm Sr_{2}RuO_{4}}$~\cite{rf:NoTOPT},
it has been recognized that the electron correlation 
plays an important role in realizing their superconductivities, by treating the electron
correlation by perturbation theory based on Fermi liquid theory.

As described above, recent neutron scattering experiments done below $T_{\rm c}$ revealed
a strong relationship between the superconductivity and magnetic
excitations of antiferromagnetically  ordered moments in ${\rm UPd_{2}Al_{3}}$.
Indicated by these experimental facts, several
groups~\cite{rf:tonelriron,rf:MS,rf:SM} have proposed a new
mechanism mediated by spin wave in the ordered antiferromagnetic state.
This mechanism is called ``magnetic exciton'' mechanism.
Magnetic excitons are a kind of spin waves that ripple through a magnetically
ordered medium, but they require a certain amount of energy to be excited.
They argued that these magnetic excitons may produce an effective interaction between the 
itinerant electrons, and so be responsible for superconductivity in a manner analogous
 to that played by phonons in conventional superconductors.
Therefore, for ``magnetic exciton'' mechanism, the coupling between the ``localized
subsystem''and the ``itinerant subsystem'' is essential to deriving the
superconductivity.

The proposal seems to be interesting as a new mechanism.
However, it is not clear whether this mechanism can actually realize
the superconductivity or not, since there is no quantitative calculation
of the critical temperature $T_{\rm c}$.
To the present authors, it 
seems to be natural that the largest Fermi sheet with large electron
mass itself leads to the superconductivity
 by a mechanism due to the effective interaction
between electrons in the ``itinerant subsystem''.

In this paper, we investigate 
the possibility that  the superconductivity in ${\rm UPd_{2}Al_{3}}$
is derived from the electron correlation effects existing  in ``itinerant subsystem'',
by the perturbation theory with respect to Coulomb repulsion $U$.
We show that the superconductivity in ${\rm UPd_{2}Al_{3}}$ can be
explained by the usual spin fluctuation mechanism in the ``itinerant subsystem''.
This explanation will be given on the basis of detailed calculation where 
the dependence on the parameters inherent in ${\rm UPd_{2}Al_{3}}$ are
carefully studied.
We concentrate our attention on 
the calculation of the superconducting transition
temperature $T_{{\rm c}}$, so  we do not investigate the 
 relationship between superconductivity and magnetic 
 excitation of antiferromagnetically 
ordered moments in ${\rm UPd_{2}Al_{3}}$ far below $T_{{\rm c}}$, in this paper.

The organization of this paper is as follows. In $\S 2$ formulation is
presented, including model Hamiltonian, perturbation expansion terms and
${\rm \acute{E}}$liashberg's equation. In $\S 3$ calculation results of
$T_{\rm c}$ and other various quantities are shown. Finally in $\S 4$
discussions and conclusion are presented.
\section{Formulation }
\subsection{Model and Hamiltonian}
\begin{figure}
\includegraphics[width=0.5 \linewidth]{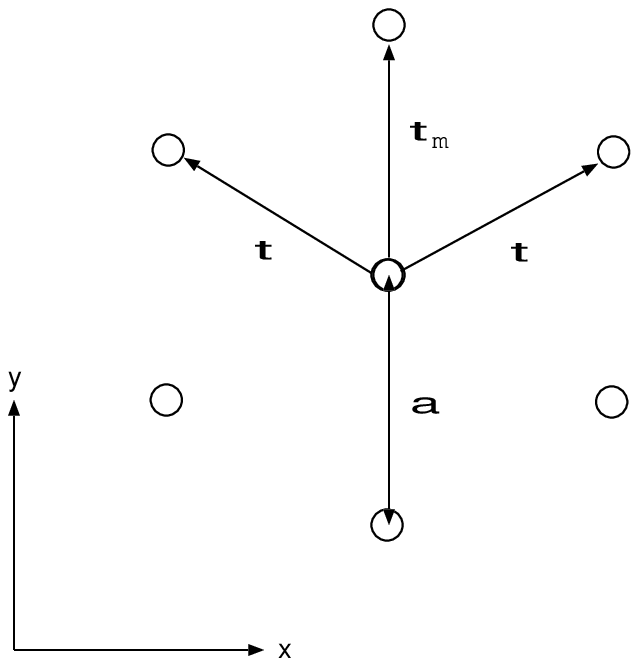}
\caption{Schematic figure of the hexagonal basal plane of ${\rm UPd_{2}Al_{3}}$. Circles in this
figure represent U atoms. }
\label{fig:latreal}
\end{figure}

Taking the importance of the correlation between
electrons in the ``itinerant subsystem'' into account
in realizing the d-wave superconductivity,
we consider only a single band of ${\rm UPd_{2}Al_{3}}$ possessing the
largest effective mass.
According to the  band calculation~\cite{rf:band3ref}, the band of ${\rm
UPd_{2}Al_{3}}$ possessing the largest
effective mass (which is the so called ``party hat'') 
has the property of quasi-two-dimensional. Therefore we 
represent the band as a  simple effective 
two-dimensional Hubbard model on the triangular lattice.
We set $x$-$y$ plane in the hexagonal basal plane, and set $y$-axis
parallel to antiferromagnetic moments. We consider only the nearest neighbor
hopping integrals and we assume for the  $y$-direction hopping integral
a different value $t_{\rm m}$ from other
hopping integrals $t$,
because the superconductivity of ${\rm UPd_{2}Al_{3}}$ is realized in
the antiferromagnetic state~\cite{rf:magstruc}(see Fig.~\ref{fig:latreal}).
So we consider that the effect of the antiferromagnetic
order(the effect of the ``localized subsystem'') is
included in the difference between $t_{\rm m}$ and $t$, and we determine
the values so as to reproduce the considered Fermi sheet which is
obtained by the band calculation and is not of hexagonal
symmetry, reflecting the antiferromagnetic structure(``localized subsystem'').

We rescale length, energy, temperature, time by
$a, t, \frac{t}{k_{\rm B}}, \frac{\hbar}{t}$ respectively
(where $a, k_{\rm B}, \hbar$ are the  lattice constant of the hexagonal
basal plane, Boltzmann constant, Planck constant divided by $2\pi$
respectively),
 we write our model Hamiltonian as follows:
\begin{eqnarray}
H&=&H_{0}+H_{1},\\
H_{0}&=&\sum_{{\bf k},\sigma}\left(\epsilon
({\bf k})-\mu
\right)
a_{{\bf k}\sigma}^{\dagger}a_{{\bf k}\sigma},
\end{eqnarray}
\begin{equation}
\epsilon({\bf k})=-4\cos(\frac{\sqrt{3}}{2}k_{x})\cos(\frac{1}{2}k_{y}
)-2t_{\rm m}\cos(k_{y}),
\end{equation}
\begin{equation}
H_{1}=\frac{U}{2N}
\sum_{\sigma\neq\sigma^{\prime}}
\sum_{{\bf k}_{1}{\bf k}_{2}{\bf k}_{3}
{\bf k}_{4}}
\delta_{{\bf k}_{1}+{\bf k}_{2},
{\bf k}_{3}+{\bf k}_{4}}
a_{{\bf k}_{1}\sigma}^{\dagger}
a_{{\bf k}_{2}\sigma^{\prime}}^{\dagger}
a_{{\bf k}_{3}\sigma^{\prime}}a_{{\bf k}_{4}\sigma},
\end{equation}
where $a_{{\bf k}\sigma}^{\dagger}(a_{{\bf k}\sigma})$
is the creation(annihilation) operator for the electron with
momentum ${\bf k}$ and spin index $\sigma$; 
$t_{\rm m}$ and $\mu$ are 
the $y$-direction hopping integral and the chemical potential, 
respectively. The sum over ${\bf k}$ indicates taking summation over a
primitive cell of the inverse lattice.
We set this  primitive cell of the inverse lattice as shown in Fig.~\ref{fig:invlat}
for the convenience of our numerical calculation.

\subsubsection{model parameters}
\begin{figure}
\includegraphics[width=0.5 \linewidth]{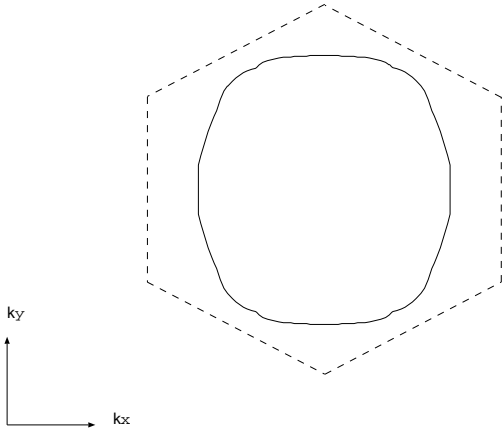}
 \caption{The bare Fermi surface calculated by our model. The region enclosed by the dashed line
 is the first Brillouin zone. See also Fig.5(b) in ref. ~\citen{rf:band3ref}.}
\label{fig:bareFSbest}
\end{figure}

Our model parameters are temperature $T$, $y$-direction hopping integral
$t_{\rm m}$, the Coulomb repulsion $U$ and the electron number $n$ per one spin site.
The bare bandwidth $\Delta\epsilon(t_{\rm m})$ of our model is
\begin{equation}
\Delta\epsilon(t_{\rm m})=
\left\{
\begin{array}{@{\,}ll}
8&\mbox{($0<t_{\rm m}<0.5$)}.\\
4+4t_{\rm m}+1/t_{\rm m}&\mbox{($t_{\rm m}\ge 0.5$)}.
\end{array}
\right.
\end{equation}
According to the band calculation and 
the de Haas-van Alphen  experiment~\cite{rf:band3ref},the parameter region of our model  
 which reproduces the Fermi sheet is given by
$0.50\le t_{\rm m}\le 0.75$ and $0.51\le n\le 0.572.$ In this parameter
region, the range of the bare bandwidth is $8\le\Delta\epsilon(t_{\rm m})\le
8.333\cdots $. We consider that $t_{\rm m}=0.75, n=0.572$ are the best
fitting values of 
parameters which well reproduce the Fermi sheet we consider. In
Fig.~\ref{fig:bareFSbest}, the bare Fermi surface calculated by our model
(the model parameters are $t_{\rm m}=0.75, n=0.572, U=0 $ and $ T=0$) is presented.
From this figure we can see that the
considered Fermi sheet determined by the band calculation~\cite{rf:band3ref}
 is well reproduced within our effective model.

The superconducting transition temperature determined
by experiments is  about 2K~\cite{rf:tyoudenAFexp}.Then the rescaled superconducting transition
temperature $T_{\rm c}$ is $T_{\rm c}\simeq 1.4\times
10^{-3}$, when we set $1$eV as the value of bandwidth obtained 
by the band calculation~\cite{rf:band3ref,rf:yamaga}.
\subsection{Green's functions}
\subsubsection{Bare Green's functions}
The bare Green's function is the following,
\begin{equation}
G^{(0)}(k)
=\frac{1}{{\rm i}\epsilon_{n}-\epsilon({\bf k})+\mu},
\end{equation}
where $\epsilon_{n}=(2n+1)\pi T(n:{\rm integer})$ is the fermion-Matsubara
frequency and short notation
$k=({\bf k},\epsilon_{n})$ is adopted.
We consider the Hartree term is included in the chemical potential.
\subsubsection{Dressed normal Green's functions}
\begin{figure}
\includegraphics[width=0.5 \linewidth]{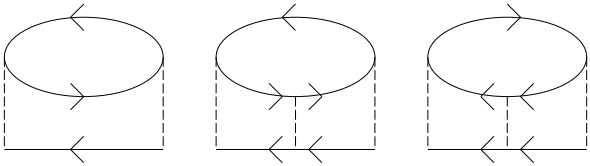}
\caption{The Feynman diagrams of the normal self-energy up to third
 order. The solid and dashed lines correspond to the bare Green's function 
 and the interaction, respectively.}
\label{fig:normal3rd}
\end{figure}

Next we consider the dressed normal Green's function.
When we consider the situation near the superconducting transition temperature,
the Dyson-Gorkov's equation can be linearized. Therefore 
the dressed normal Green's function is obtained from the bare Green's
function with only the  normal self-energy correction.
We expand the normal self-energy  up to third order with
respect to $U$;the diagrams are shown in Fig.~\ref{fig:normal3rd}.

Then we obtain the normal self-energy as follows,
%\begin{full}
\begin{equation}
\Sigma_n(k)=
\frac{T}{N}\sum_{k^{\prime}}
[U^{2}\chi_0(k- k^{\prime})
+U^{3}\chi_0^{2}(k- k^{\prime})
+U^{3}\phi_0^{2}(k+k^{\prime})
]G_0(k^{\prime}),
\end{equation}
%\end{full}

where $\chi_0(\cdots)$ and $\phi_0(\cdots)$ are given respectively as
\begin{equation}
\chi_0({\bf q},\omega_m)=-\frac{T}{N}\sum_{{\bf k},n}
G_0({\bf k},\epsilon_n)G_0({\bf q}+{\bf k},\omega_m+\epsilon_n),
\end{equation}
\begin{equation}
\phi_0({\bf q},\omega_m)=-\frac{T}{N}\sum_{{\bf k},n}
G_0({\bf k},\epsilon_n)G_0({\bf q}-{\bf k},\omega_m-\epsilon_n).
\end{equation}

Here $\omega_{m}=2m\pi T\nonumber(m:{\rm integer})$
 is the boson-Matsubara frequency. The quantity $\chi_{0}({\bf q},\omega_{m})$ has
the physical meaning of the bare susceptibility and expresses spin
fluctuations in the system. More over, the bare susceptibility plays an
important role in the calculation of $T_{\rm c}$, that is, it determines the 
magnitude and the spatial and temporal variation of the effective
interaction between electrons in the ``itinerant subsystem'',
through the higher order terms in  $U$. (See the first term of right hand side
 of the equation 2.13.)

Note that the Hartree term has been already included in the chemical potential
and the constant terms which have not been included in the Hartree term  are
included in the chemical potential shift when we fix the particle
number.

Then the dressed normal Green's function is 
\begin{equation}
G(k)=\frac{1}{{\rm i}\epsilon_n-(\epsilon({\bf k})-\mu-\delta\mu+\Sigma_n(k))},
\end{equation}
where $\delta\mu$ is determined so that the following equation is satisfied.
\begin{equation}
n=\frac{T}{N}\sum_{k}G(k)=\frac{T}{N}\sum_{k}G_0(k).
\end{equation}
We expand the above equation up to the third order
of the interaction with regard to $\delta\mu-\Sigma_n(k)$, we obtained  $\delta\mu$ as 
\begin{equation}
\delta\mu=-\frac{\frac{T}{N}\sum_{k}G_0^{2}(k)\Sigma_n(k)}{\chi_0({\bf 0},0)}.
\end{equation}

\subsubsection{Anomalous self-energy and effective interaction}
\begin{figure}
\includegraphics[width=0.5 \linewidth]{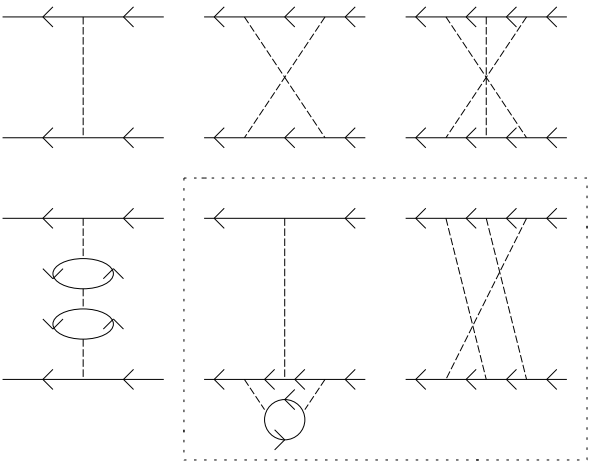}
\caption{The Feynman diagrams of the effective interaction up to third
 order. The solid and dashed lines correspond to the bare Green's function 
 and the interaction, respectively. The diagrams enclosed by the dashed
 line are vertex corrections. The other diagrams are included in RPA.}
\label{fig:EIsimpl3rd}
\end{figure}

When we consider the situation
near the superconducting transition temperature, the anomalous self-energy $\Sigma_{a}(k)$ is 
represented by the anomalous Green's function $F(k)$ and the (normal)effective
interaction. We expand the effective interaction up to third order with
respect to $U$ as shown in Fig.~\ref{fig:EIsimpl3rd}.

Then we obtain the anomalous self-energy as follows;
%\begin{full}
\begin{eqnarray}
\Sigma_a(k)&=&
-\frac{T}{N}\sum_{k^{\prime}}[
U+U^{2}\chi_0(k+k^{\prime})+2U^{3}\chi_0^{2}(k+k^{\prime})
]F(k^{\prime})\nonumber\\
& &\mbox{}-U^{3}\frac{T^{2}}{N^{2}}\sum_{k^{\prime}k^{\prime\prime}}
G_0(k^{\prime})[\chi_0(k+k^{\prime})-\phi_0(k+k^{\prime})]
G_0(k+k^{\prime}-k^{\prime\prime})F(k^{\prime\prime})
\nonumber\\
& &\mbox{}
-U^{3}\frac{T^{2}}{N^{2}}
\sum_{k^{\prime}k^{\prime\prime}}
G_0(k^{\prime})[\chi_0(-k+k^{\prime})-\phi_0(-k+k^{\prime})]G_0(-k+k^{\prime}-k^{\prime
\prime})F(k^{\prime\prime}).
\end{eqnarray}
%\end{full}

\subsection{$\acute{E}$liashberg's equation}
From the linearized Dyson-Gorkov equation, we obtain the anomalous Green's
 function as follows;
\begin{equation}
F(k)=|G(k)|^{2}\Sigma_a(k).
\end{equation}
Then the ${\rm\acute{E}liashberg}$'s equation is given by

%\begin{full}
\begin{eqnarray}
\Sigma_a(k)&=&-\frac{T}{N}\sum_{k^{\prime}}[
U+U^{2}\chi_0(k+k^{\prime})+2U^{3}\chi_0^{2}(k+k^{\prime})
]|G(k^{\prime})|^{2}\Sigma_a(k^{\prime})\nonumber\\
& &\mbox{} -U^{3}\frac{T^{2}}{N^{2}}\sum_{k^{\prime}k^{\prime\prime}}
G_0(k^{\prime})[\chi_0(k+k^{\prime})-\phi_0(k+k^{\prime})]G_0(k+k^{\prime}-k^{\prime\prime})|
G(k^{\prime\prime})|^{2}
\Sigma_a(k^{\prime\prime})
\nonumber\\
& &\mbox{}
-U^{3}\frac{T^{2}}{N^{2}}
\sum_{k^{\prime}k^{\prime\prime}}
G_0(k^{\prime})[\chi_0(-k+k^{\prime})-\phi_0(-k+k^{\prime})]G_0(-k+k^{\prime}-k^{\prime\prime})|G(k^{\prime\prime}
)|^{2}\Sigma_a(k^{\prime\prime}).\nonumber\\
\end{eqnarray}
%\end{full}

We consider that the system is superconducting state when the eigen value 
of this equation is 1.
\section{Calculation Results }
\subsection{Details of the numerical calculation}
To solve the ${\rm\acute{E}}$liahberg's equation by using the
power method algorithm, we have to calculate the 
summation over the momentum and the frequency space. Since all summations
 are in the convolution forms, we can carry out them by using the
 algorithm of the Fast Fourier Transformation.
For the frequency, irrespective of the temperature, 
we have 1024 Matsubara
frequencies. Therefore we calculate throughout 
in the temperature region $T\ge T_{\rm lim}$
, where $T_{\rm lim}$ is the lower limit temperature
 for reliable numerical calculation,
which is estimated about $2.0\times 10^{-3}
(>\Delta\epsilon(t_{\rm m})/(2\pi\times 1024)\simeq 1.3\times 10^{-3})$; we divide a primitive cell into 128$\times$128 meshes.

We have carried out analytically continuing procedure by using  Pad${\rm\acute{e}}$
 approximation.
\subsection{Dependence of $T_{\rm c}$ on $U, t_{\rm m}$ and $n$}
To solve the ${\rm \acute{E}}$liashberg's
equation, we set the initial gap function($d_{xy}$-symmetry) as follows.
\begin{equation}
\Sigma_{a}(k)\propto \sin\frac{\sqrt{3}}{2}k_{x}\sin\frac{1}{2}k_{y}.
\end{equation}
Notice that the ``itinerant subsystem'' has no hexagonal
symmetry due to the antiferromagnetic structure(``localized subsystem'').
The calculated gap functions show the node at $k_{x}=0$ and $k_{y}=0$ and changes
the sign across the node for all parameters. The symmetry of Cooper pair is $d_{xy}$.

We calculate $T_{\rm c}$ around the best fitting values of 
parameters($t_{\rm m}=0.75,n=0.572$) because the best fitting values of 
parameters have some arbitrariness.

\begin{figure}
\includegraphics[width=1.0 \linewidth]{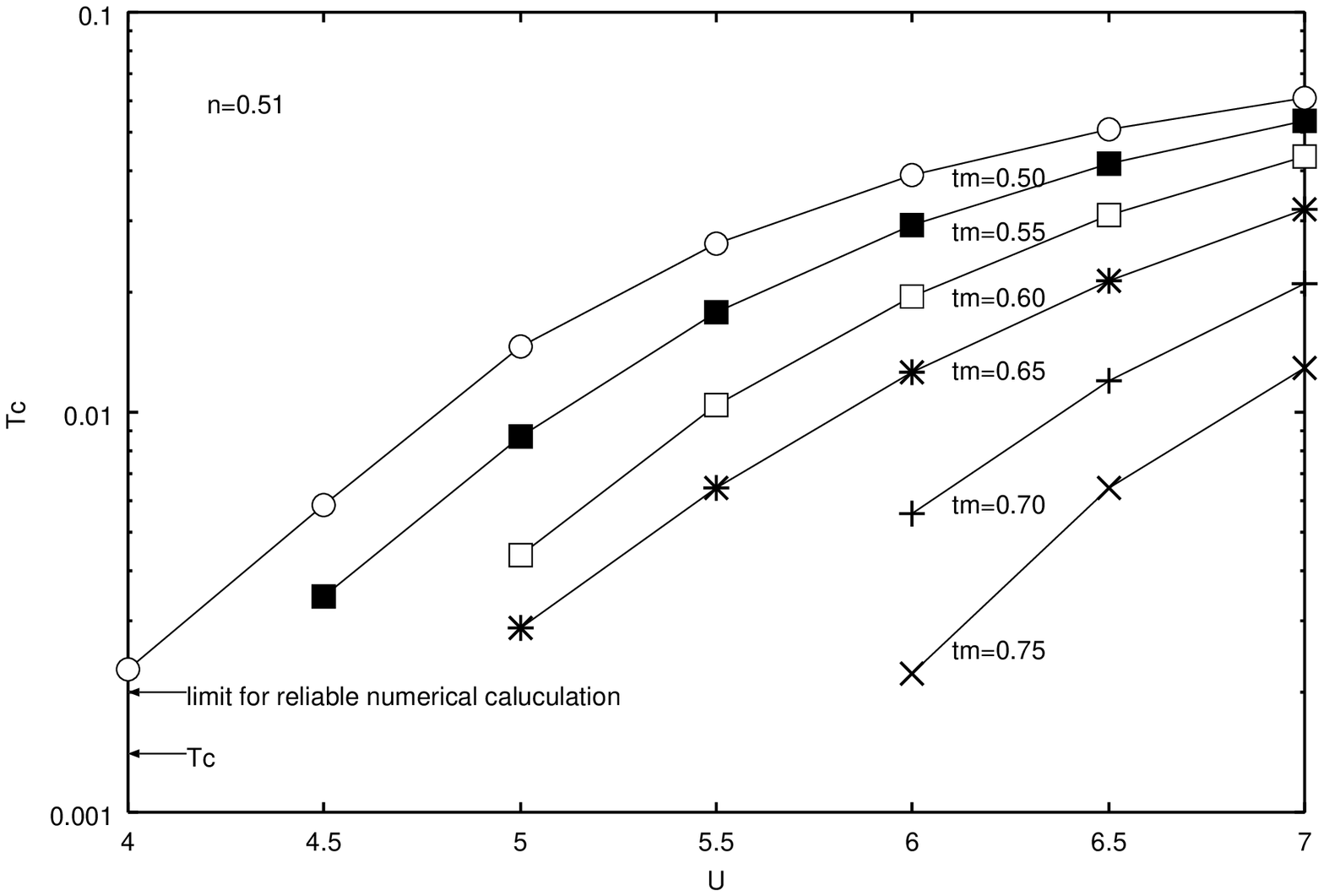}
\caption{The calculated $T_{\rm c}$ as $U$ is varied, at $n=0.51$ and for various 
 values of $t_{\rm m}$ as shown in the figure.}
\label{fig:nTc0.51}
\end{figure}

\begin{figure}
\includegraphics[width=1.0 \linewidth]{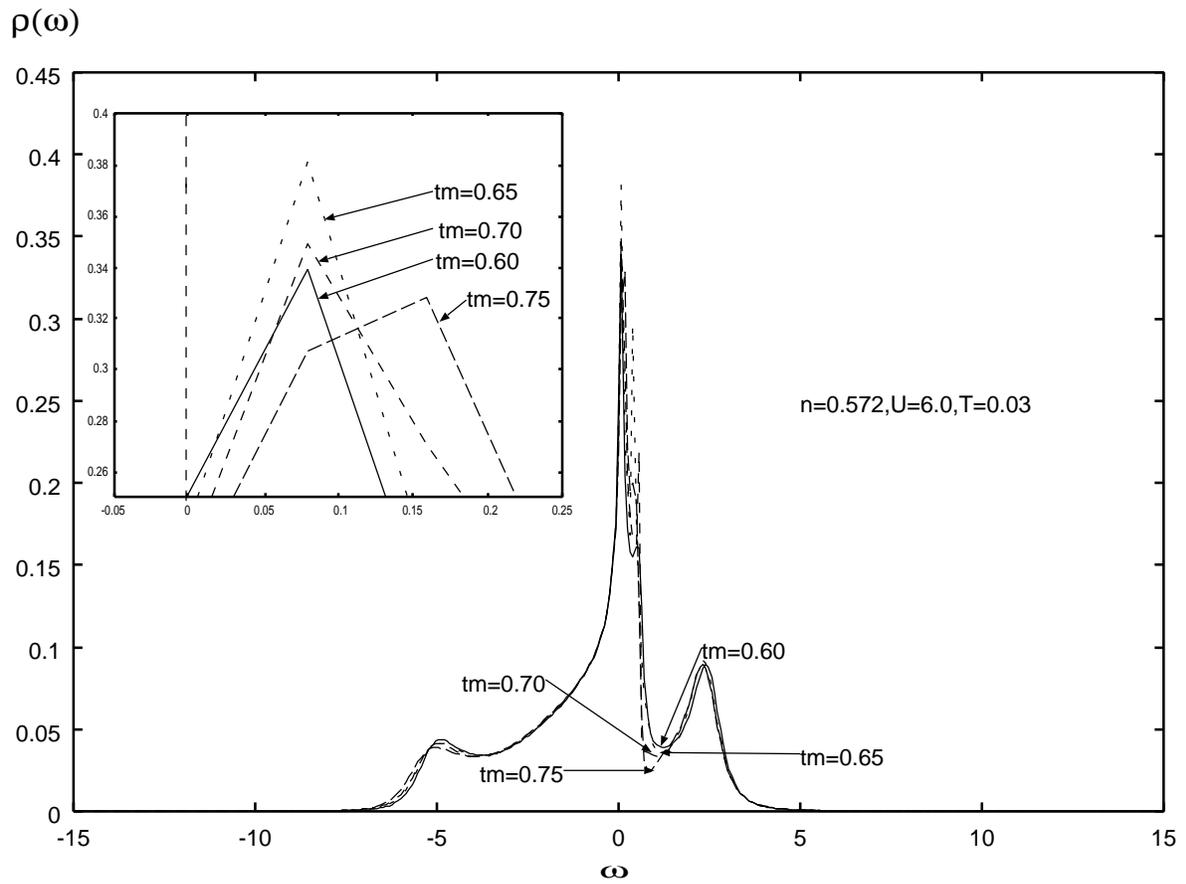}
\caption{The density of states as $t_{\rm m}$ is varied, at
 $n=0.572$, $U=6.0$ and $T=0.03$. The inset shows the details near 
 the Fermi level.}
\label{fig:DOStmn0.572u6.0t0.03}
\end{figure}

\begin{figure}
\includegraphics[width=1.0 \linewidth]{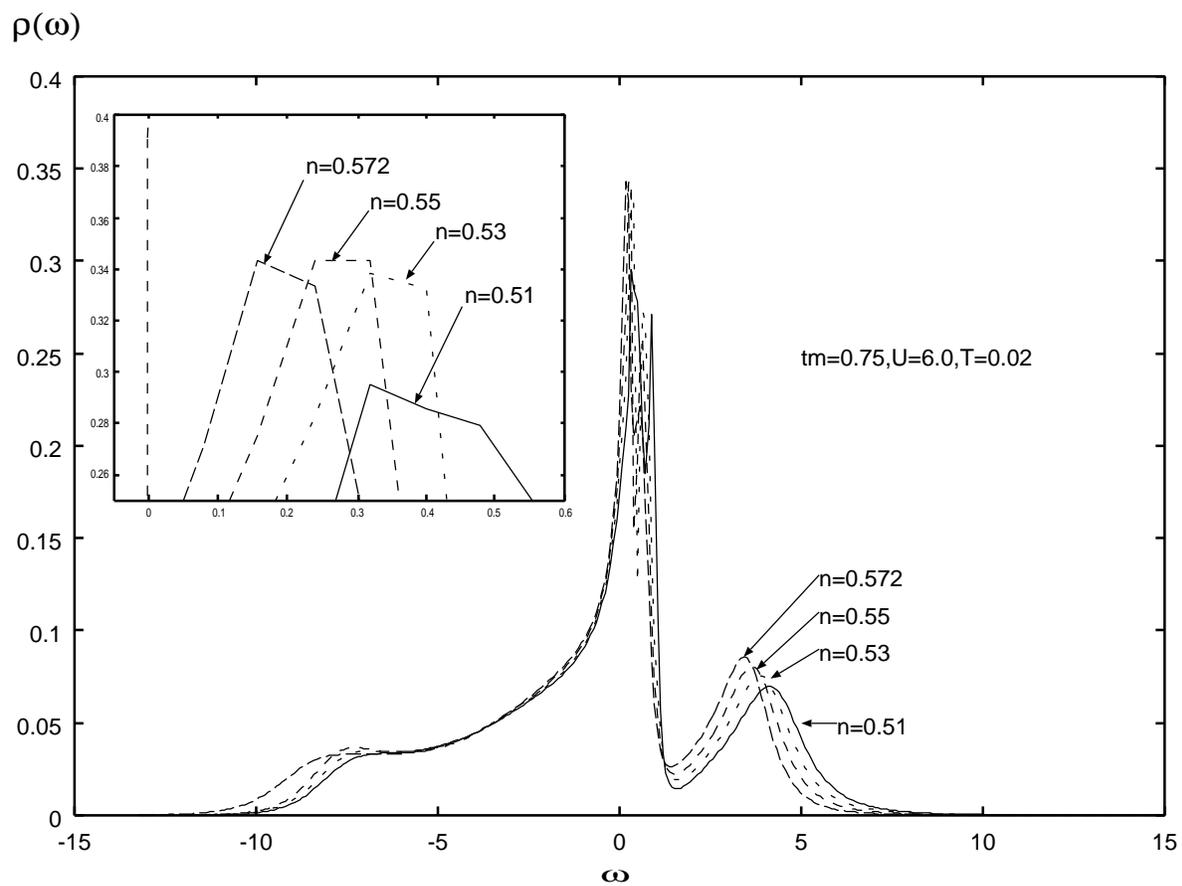}
\caption{The density of states as $n$ is varied,at $t_{\rm m}=0.75$, $U=6.0$
 and $T=0.02$. The inset shows the details near 
 the Fermi level.}
\label{fig:DOSntm0.75u6.0t0.02}
\end{figure}

The dependence of $T_{\rm c}$ on $U, n$ and $t_{\rm m}$ are shown in
Figs.~\ref{fig:n0.572Tc}$\sim$~\ref{fig:nTc0.51}.
From these results, we can point out the following
facts. For large $U$ higher $T_{\rm c}$ are
obtained commonly for all parameters.
 When we fix the electron number $n$ per one spin site 
and the Coulomb repulsion $U$, for large $t_{\rm m}$, the system get close to
real triangular lattice ($t_{{\rm m}}=1$) for electrons in the ``itinerant subsystem'' and at the same time
 Fermi level goes away from the van Hove singularity(see Fig.~\ref{fig:DOStmn0.572u6.0t0.03} shown later), then lower $T_{\rm c}$ are obtained.
Contrary, when we fix the $y-$direction hopping integral
$t_{\rm m}$ and the Coulomb repulsion $U$, for large $n$, the system becomes away from the
half-filling state ($n=0.5$) and at the same time Fermi level gets close to the
van Hove singularity(see Fig.~\ref{fig:DOSntm0.75u6.0t0.02} shown later), then  higher
$T_{\rm c}$ are obtained.

\subsection{Vertex corrections}
\begin{figure}
\includegraphics[width=1.0 \linewidth]{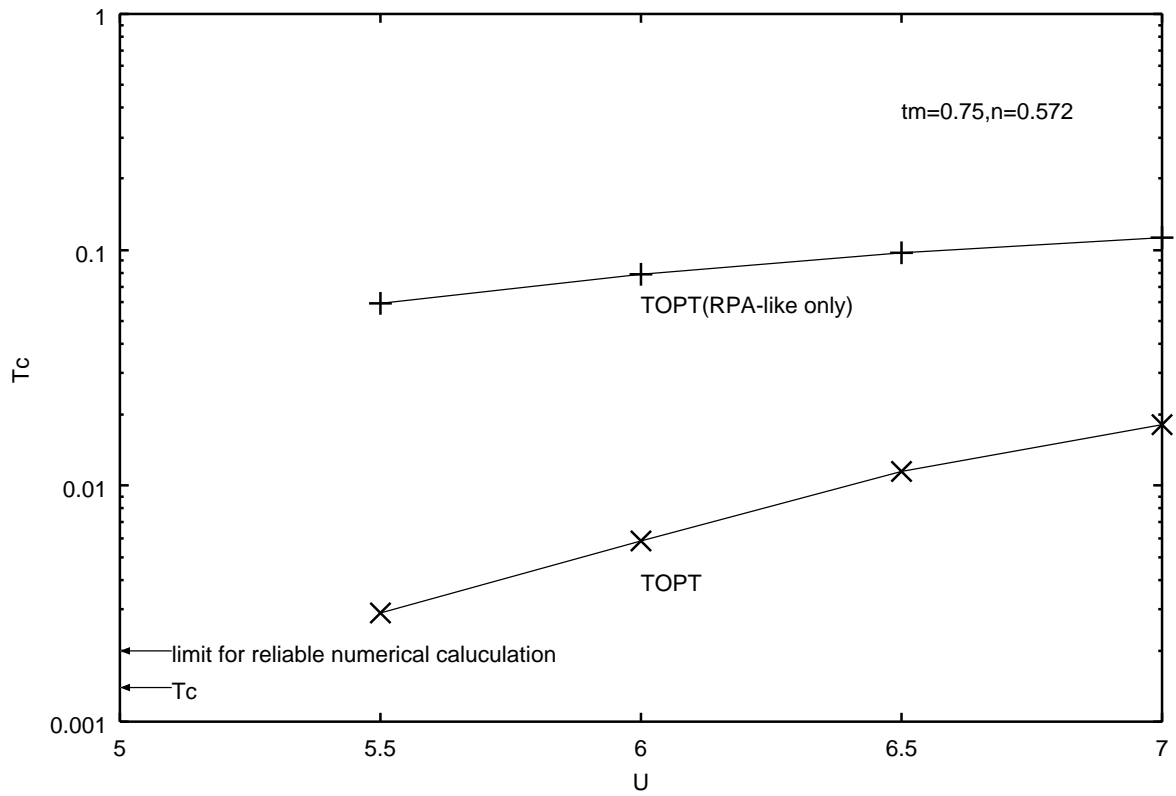}
\caption{The calculated $T_{\rm c}$. TOPT and TOPT(RPA-like only) in this figure mean
 that full diagrams and only RPA-like diagrams of anomalous self-energies up to third
 order are included, respectively. See also Fig. 4.}
\label{fig:RPA-TOPT}
\end{figure}

To examine how the vertex corrections influence  $T_{\rm c}$,
we calculate $T_{\rm c}$ by including only RPA-like
 diagrams of anomalous self-energies up to third order, in other
 words, without the vertex corrections and 
compare obtained $T_{\rm c}$ with $T_{\rm c}$ calculated by including full diagrams of
anomalous self-energies up to third order(Fig.~\ref{fig:RPA-TOPT}).
From this figure, we can see that $T_{\rm c}$ calculated by including only RPA-like
 diagrams of anomalous self-energies up to third order is higher than 
$T_{\rm c}$ calculated by including full diagrams of
them. 
This figure shows that the main origin of the d-wave superconductivity
is the momentum and frequency dependence of  
spin fluctuations given by the RPA-like terms;
 spin fluctuations exist in our
nested Fermi sheet, and 
the Coulomb interaction among electrons in ``itinerant subsystem''.
The vertex corrections reduce $T_{\rm c}$ by one order of magnitude.
So the vertex corrections is important for obtaining reasonable $T_{\rm c}$.

\subsection{Behavior of $\chi_{0}(q)$}
\begin{figure}
\includegraphics[width=1.0 \linewidth]{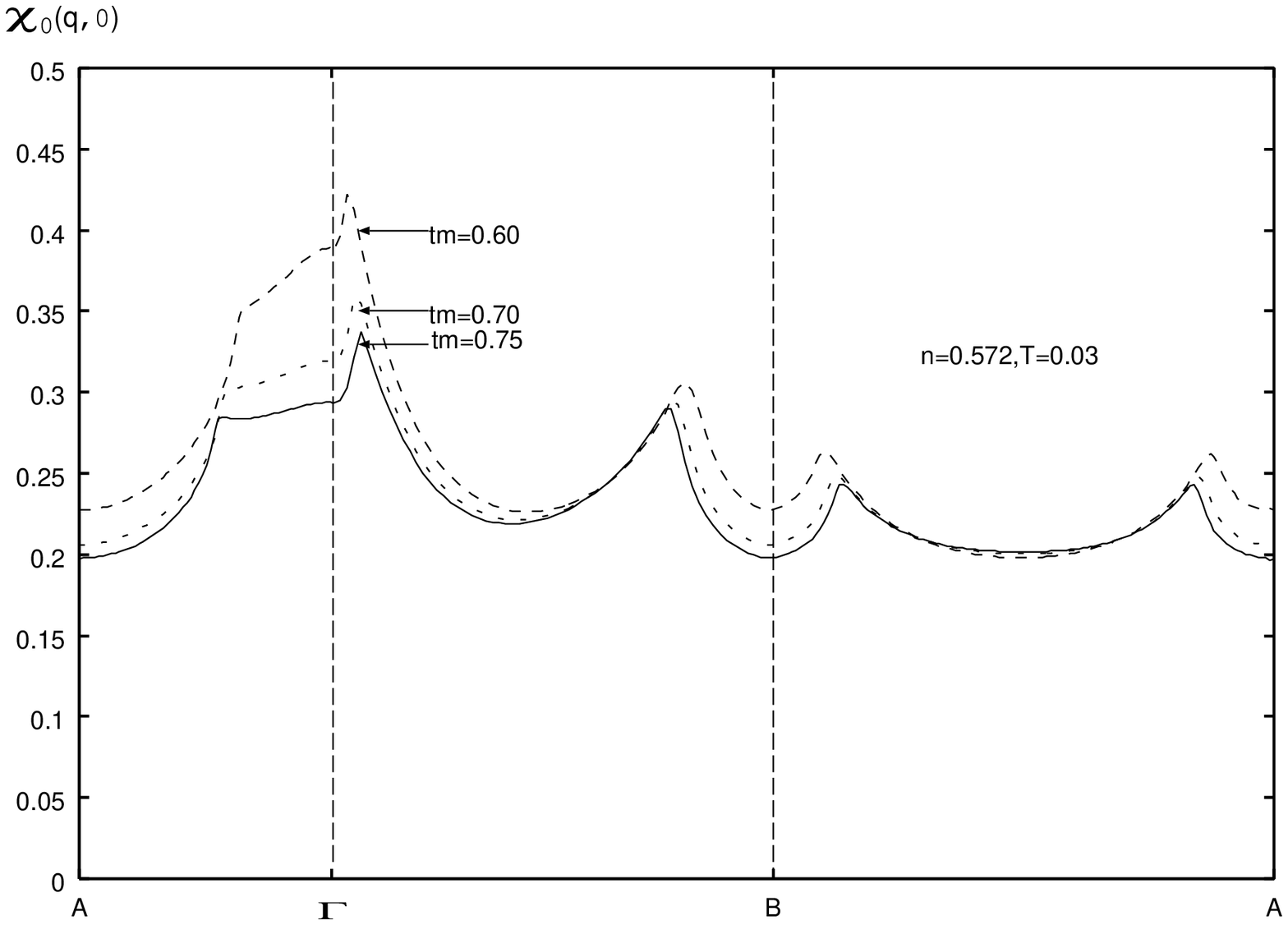}
\caption{The momentum dependence of the static bare susceptibility for
 various $t_{\rm m}$. The notations under the horizontal axis are defined in
 Fig.2. These results are obtained for $n=0.572$, $T=0.03$.}
\label{fig:X0tmn0.572t0.03}
\end{figure}

\begin{figure}
\includegraphics[width=1.0 \linewidth]{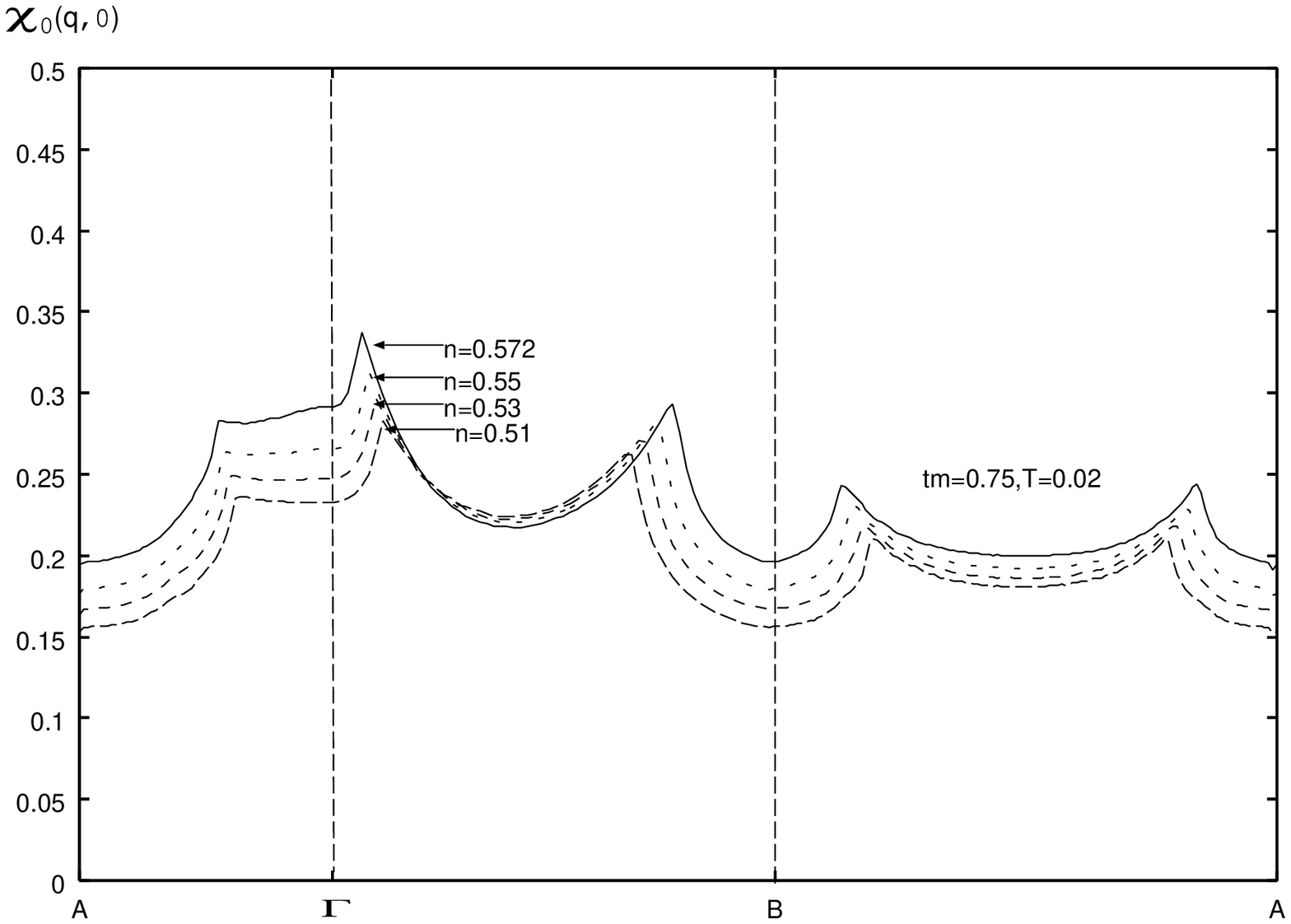}
\caption{The momentum dependence of the static bare susceptibility for
 various $n$. The notations under the horizontal axis are defined in
 Fig.2. These results are obtained for $t_{\rm m}=0.75$, $T=0.02$.}
\label{fig:X0ntm0.75t0.02}
\end{figure}

The calculated results of the static bare susceptibility are shown in
Fig.~\ref{fig:X0tmn0.572t0.03}$ \sim$~\ref{fig:X0ntm0.75t0.02} for
various value of $t_{\rm m}$ and $ n$.
From these figures, we point out the following facts.
For all parameters, $\chi_{0}({\bf q},0)$ have sufficiently strong
momentum dependence, although there exist no prominent peak.
When we fix the electron number $n$ per one spin site
and increase the parameter $t_{\rm m}$, the system approaches the real triangular lattice for
electrons in the ``itinerant subsystem''. In this case the peak and 
the momentum dependence 
of $\chi_{0}({\bf q},0)$ is slightly
suppressed since the antiferromagnetic spin fluctuations are suppressed
by the frustration effect in triangular lattice.
Contrary, when we fix the $y-$direction hopping integral $t_{\rm m}$ and
increase the electron number $n$ per one spin site, 
the system become away from the half-filling state and
the peak and the 
momentum dependence of $\chi_{0}({\bf q},0)$ is slightly enhanced.
The reason is that Fermi surface possesses some nesting properties with
increasing $n$ from half-filling case, where the Fermi surface is similar 
to a circle without nesting properties.

\subsection{Density of states}
The density of states(DOS) is given by 
\begin{equation}
\rho(\omega)=-\frac{1}{N\pi}\sum_{{\bf k}}{\rm Im}G^{R}({\bf k},\omega),
\end{equation}
where
\begin{displaymath}
G^{R}({\bf k},\omega)=\left.G({\bf k},\epsilon_{n})\right|_{{\rm i}\epsilon_{n}\rightarrow \omega+{\rm i}\eta}.
\end{displaymath}
We show
the $t_{\rm m}$ and $n$-
dependence of DOS in Figs.~\ref{fig:DOStmn0.572u6.0t0.03} and ~\ref{fig:DOSntm0.75u6.0t0.02}.
From insets in these figures,
we can see that the position of the van Hove singularity shifts
upward from the Fermi
level with increasing $t_{{\rm m}}$ and decreasing $n$.
This departure of the van Hove singularity from the Fermi level reduces
the superconducting transition temperature $T_{{\rm c}}$.

\subsection{Self-energy}
The self-energy is given by
\begin{displaymath}
\Sigma_{n}^{R}({\bf k},\omega)=\left.\Sigma_{n}({\bf k},\epsilon_{n})\right|_{{\rm i}\epsilon
_{n}\rightarrow\omega+{\rm i}\eta}.
\end{displaymath}

\begin{figure}
\includegraphics[width=1.0 \linewidth]{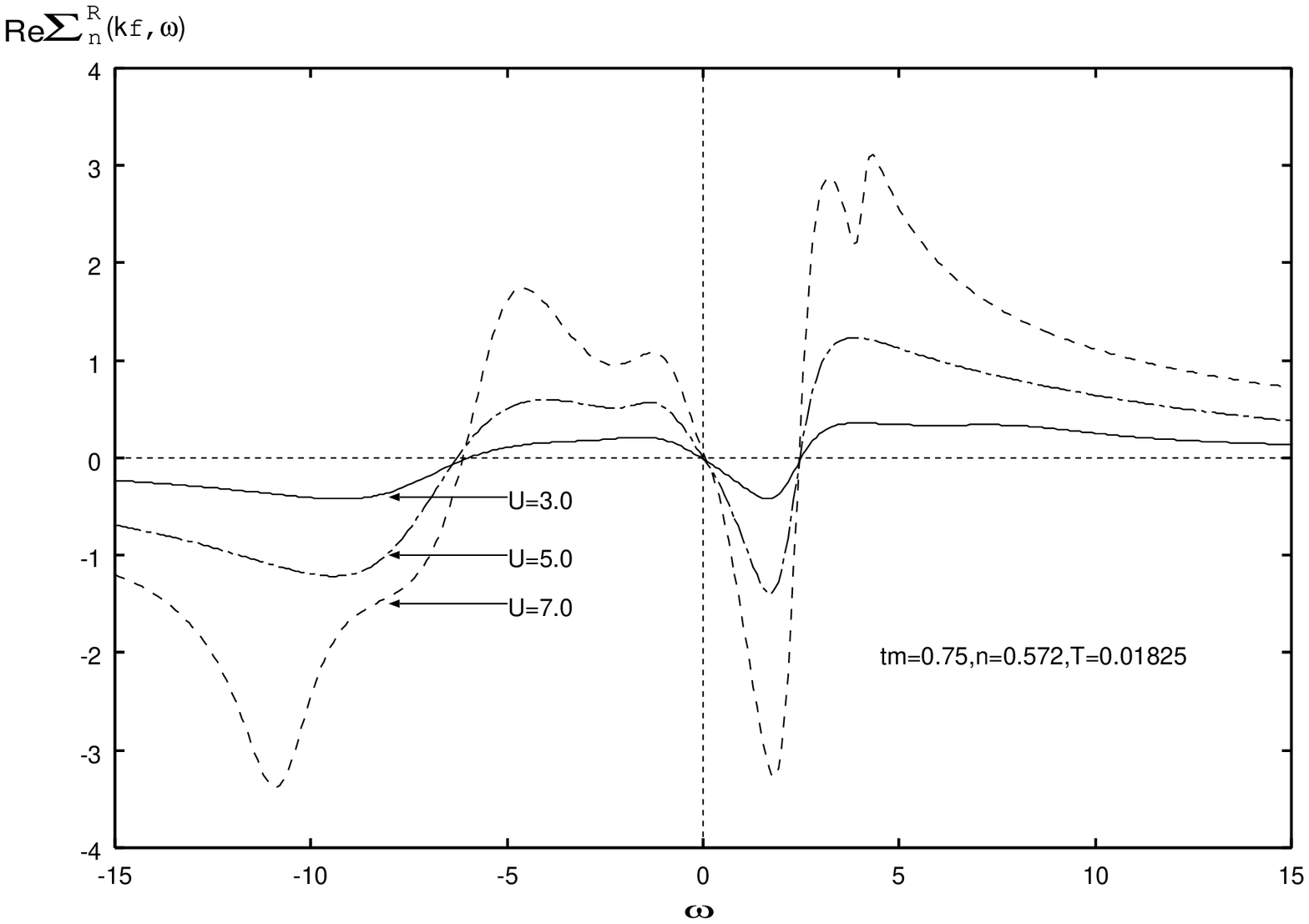}
\caption{The real part of the normal self-energy at the Fermi momentum,
 at $t_{\rm m}=0.75$, $n=0.572$, $T=0.01825$ and for various values of $U$ as shown in the figure.}
\label{fig:ReSE}
\end{figure}

\begin{figure}
\includegraphics[width=1.0 \linewidth]{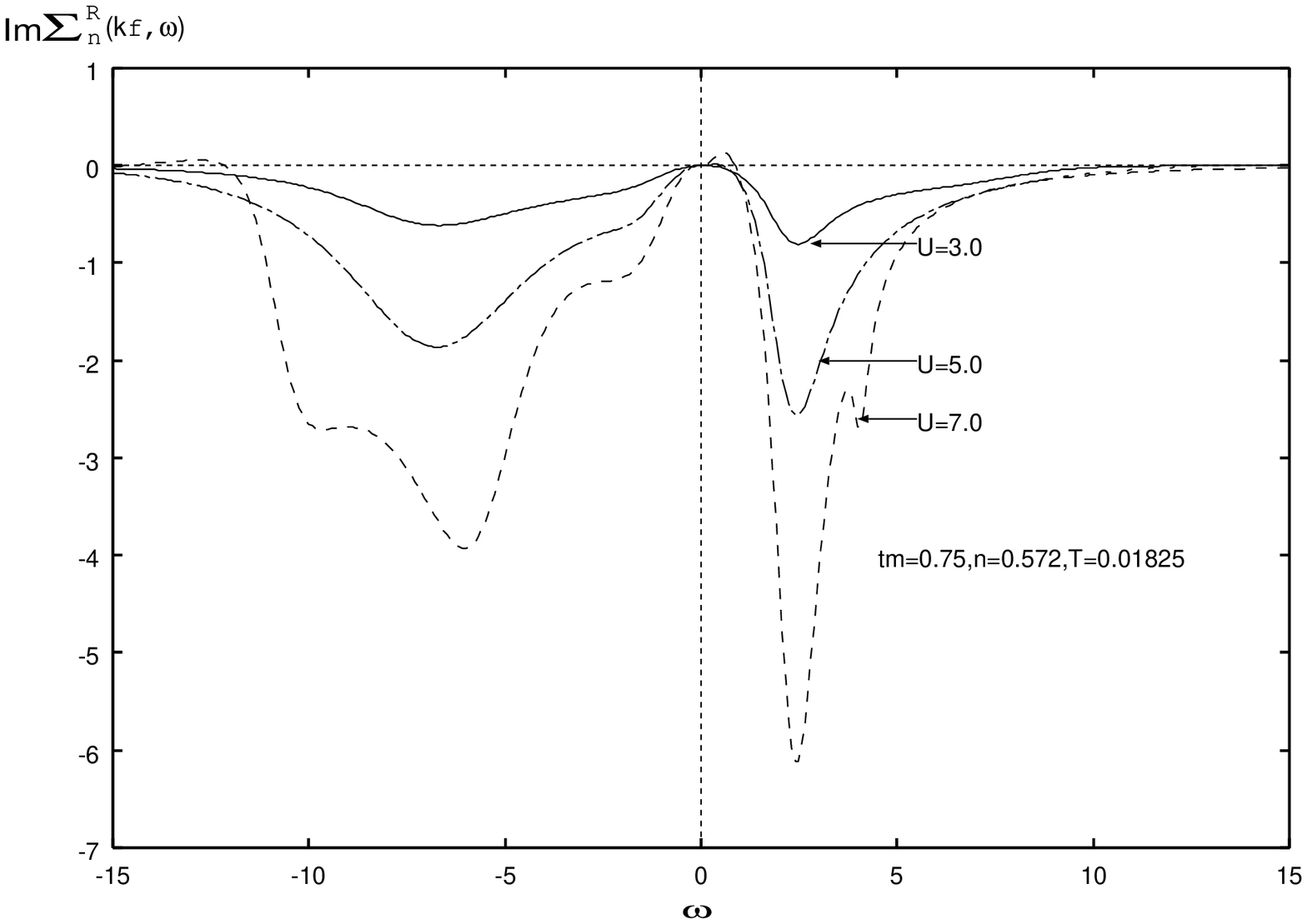}
\caption{The imaginary part of the normal self-energy at the Fermi momentum,
 at $t_{\rm m}=0.75$, $n=0.572$, $T=0.01825$ and for various values of $U$ as
 shown in the figure.}
\label{fig:ImSE}
\end{figure}

The real part and the imaginary part of the self-energy at Fermi
momentum are shown in Fig.~\ref{fig:ReSE} and Fig.~\ref{fig:ImSE} respectively.
The $\omega$-dependence of both parts near $\omega=0$ are respectively 
given by ${\rm Re}\Sigma_{n}^{R}({\bf k}_{f},\omega)\propto -\omega$ and ${\rm
Im}\Sigma_{n}^{R}({\bf k}_{f},\omega)\propto -\omega^{2}$.
This behavior is the same as that for the usual Fermi liquid.
As $U$ increases, the slope of ${\rm Re}\Sigma_{n}^{R}({\bf
k}_{f},\omega)$ at $\omega=0$ becomes steeper and the coefficient of the
$\omega^{2}$-term in ${\rm Im}\Sigma_{n}^{R}({\bf k}_{f},\omega)$ become
larger. This indicates that the mass and the damping rate of the
quasi-particle become larger as $U$ increases.
These results are the typical Fermi liquid ones.

\section{Summary, Discussion and Conclusion}
In this paper, taking the importance of the 
correlation effect between electrons 
in ``itinerant subsystem'' into account in realizing the superconductivity,
we have considered a single band of ${\rm UPd_{2}Al_{3}}$ whose
effective mass is the largest. We have represented the band with an effective 
two-dimensional Hubbard model on triangular lattice and calculated
$T_{\rm c}$ on the basis of third order perturbation theory.
Reasonable transition temperatures have been obtained for moderately large $U$.
We also have calculated $T_{\rm c}$ by including only RPA-like
 diagrams of anomalous self-energies up to third order, in other
 words, without the vertex corrections and compared it
 with $T_{\rm c}$ calculated by including full diagrams of anomalous self-energies up to third order.
We point out that 
the main origin of the superconductivity can be ascribed to 
the momentum and frequency dependence of
the spin
fluctuations and the correlation between electrons in
the ``itinerant subsystem'', although
the vertex correction terms are important for 
reducing $T_{\rm c}$ and obtaining reasonable transition temperatures.
We have calculated DOS and the normal self-energy.
The obtained behavior of both quantities is the expected
one for the Fermi liquid.

Now, we briefly discuss the  theories suggesting that the superconductivity in 
${\rm UPd_{2}Al_{3}}$ is derived from the mechanism of exchanging ``magnetic
excitons''.
Huth {\it et al.}~\cite{rf:tonelriron}
 predicted that line nodes are at the rim of the magnetic
Brillouin zone and on the ``cigar'' Fermi 
sheet. 
Their theory is based on the assumption of 
the pairing interaction mediated by the exchanging 
``magnetic excitons'' and the 
band structure calculated by K.Kn${\rm\ddot{o}pfle}$ 
{\it et al.}~\cite{rf:band2}.
Miyake and Sato~\cite{rf:MS} also predicted that the line node is on the plane
very close to the zone boundary of the folded Brillouin zone in the 
antiferromagnetically ordered state on the basis of the itinerant-localized
duality model; their theory is also based on 
 the observed behavior of the dynamical magnetic susceptibility
and the assumption of the pairing interaction induced by the exchanging 
``magnetic excitons''
but without using any information about the Fermi sheets.
According to these analysis~\cite{rf:tonelriron,rf:MS}, Fermi sheets which have
the dispersion along the crystallographic c-axis and cross the 
zone boundary plane of the folded Brillouin zone in the 
antiferromagnetically ordered state are necessary to possess the line
 nodes actually. 
On the other hand, according to the band 
calculation performed by Inada {\it et al.}~\cite{rf:band3ref}, 
the ``cigar'' Fermi sheet dose not cross 
the zone boundary plane of the folded Brillouin zone in the 
antiferromagnetically ordered state, while the
``cigar'' Fermi sheet calculated by Kn${\rm\ddot{o}pfle}$ 
{\it et al.}~\cite{rf:band2}
crosses this plane. 
Moreover,
Inada {\it et al.}~\cite{rf:band3ref,rf:cigarno} 
have tried to determine the position of the
line node on the actual Fermi surface by the de Haas-van Alphen
experiment. Their conclusion is that any line node in the anisotropic
energy gap does not exist on the ``cigar'' Fermi sheet and/or it is
difficult to distinguish experimentally the line node from the gapless
state created by the pair-breaking effects by the applied magnetic field.
Therefore, we now consider that the basis of ``magnetic exciton''
mechanism has not been confirmed.

These theories described above are indicated by the existence of the coupling 
between the superconductivity and the magnetic excitations of 
antiferromagnetically ordered moments in ${\rm
UPd_{2}Al_{3}}$ below $T_{\rm c}$. 
Generally speaking, the coupling between superconductivity and
magnetism exists more or less. 
Therefore it doesn't always follow that 
superconductivity is mainly derived from the magnetic mechanism
by the reason why there  exists the coupling between superconductivity and
magnetism. This is understood from the fact that 
the coupling between superconductivity and spin
fluctuations or phonons is detected in high-$T_{\rm c}$ superconductors but 
the superconductivities in high-$T_{\rm c}$ are not always mainly derived from
these spin fluctuations or phonons.

On the other hand, above $T_{\rm c}$, the behavior of the electronic specific-heat 
and the electrical resistivity are the typical
 Fermi liquid ones and anomalous behavior of these
quantities is not detected in ${\rm UPd_{2}Al_{3}}$ 
,for example, such as $\rho-\rho_{0}\propto T$ 
in high $T_{\rm c}$ superconductor.
Moreover the magnetic transition temperature($T_{\rm N}=14$K) is
sufficiently higher than the superconducting transition temperature($T_{\rm
c}=2$K). Therefore we consider that the coupling between local moments
(``localized subsystem'') and
electrons in the ``itinerant subsystem'' is not so strong,
while Sato {\it et al.} assume that this coupling is very
strong in ref.\citen{rf:SM}.

Therefore we considered that the electron correlation in the ``itinerant subsystem''
also plays an important role to derive the superconductivity in 
${\rm UPd_{2}Al_{3}}$.
Quasi-two-dimensional Fermi sheet so called
``party hat''
is  most favorable to superconductivity, because of the most strong electron
correlation in the ``itinerant subsystem''. Our proposition is that
 this sheet possessing large area and electron mass
 is essential to derive the superconductivity in ${\rm UPd_{2}Al_{3}}$,
while the Fermi sheets called ``cigar''and ``eggs''playing an essential role in realizing
the superconductivity mediated by ``magnetic exciton'' seem to be too
small to induce the superconductivity in the total system.

In the last, we also point out that results of tunneling experiments 
and our proposition don't contradict. 
Tunneling spectra have been also measured in this material by the same
group~\cite{rf:tonnelexp1,rf:tonnelexp2}.
Their conclusion is that an energy gap along the c-axis in  ${\rm UPd_{2}Al_{3}}$
is well resolved.
In the multi band system such an ${\rm UPd_{2}Al_{3}}$,
the superconducting transition, in principle, simultaneously  occurs in all 
Fermi sheets. Therefore the energy gaps open in all Fermi sheets below
$T_{\rm c}$, then an energy gap along the c-axis exists in  ${\rm UPd_{2}Al_{3}}$
since ${\rm UPd_{2}Al_{3}}$ is not a perfect
two-dimensional system.

In conclusion we 
present a new mechanism
that the superconductivity in
${\rm UPd_{2}Al_{3}}$ can be also derived from  the
correlation between electrons in the ``itinerant subsystem''
and the symmetry of the pairing state is $d_{xy}$.
In our model, the main origin of the superconductivity is 
the momentum and frequency dependence of
the spin fluctuations and 
the Coulomb interaction between electrons in the ``itinerant
subsystem''. This 
momentum and frequency dependence of
the spin fluctuations stems from the shape of our  Fermi
sheet which undergoes the symmetry-breakdown 
due to the antiferromagnetic order(``localized system'') and then possesses
nesting properties.

\section*{Acknowledgments}
One of the authors (Y.N) acknowledges T.Jujo for  
 advising on the numerical computation.
Numerical computation in this work was carried out at the Yukawa
 Institute Computer Facility.

\newpage
%%%%(Caption)%%%
%1

%2
\begin{figure}
\includegraphics[width=0.5 \linewidth]{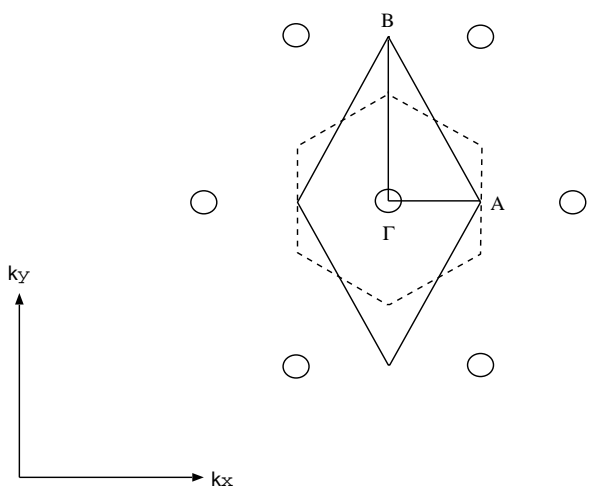}
\caption{Our primitive cell of the inverse lattice. The regions enclosed
 by the solid and dashed lines are our primitive cell and the first
 Brillouin zone, respectively. }
\label{fig:invlat.}
\end{figure}

%3

%4

%5

%6
\begin{figure}
\includegraphics[width=1.0 \linewidth]{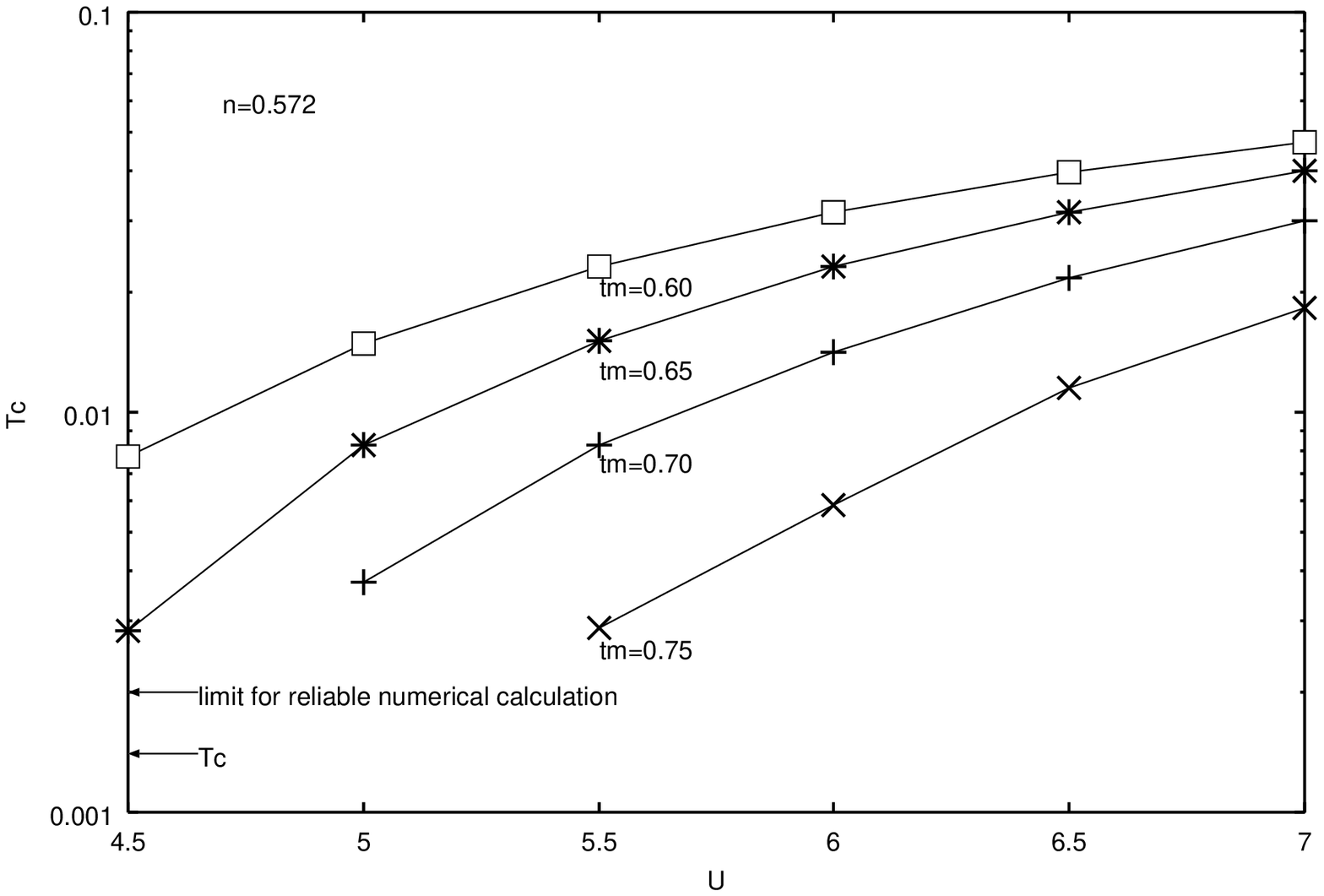}
\caption{The calculated $T_{\rm c}$ as $U$ is varied, at $n=0.572$ and for various 
 values of $t_{\rm m}$ as shown in the figure.}
\label{fig:nTc0.572}
\end{figure}

%7
\begin{figure}
\includegraphics[width=1.0 \linewidth]{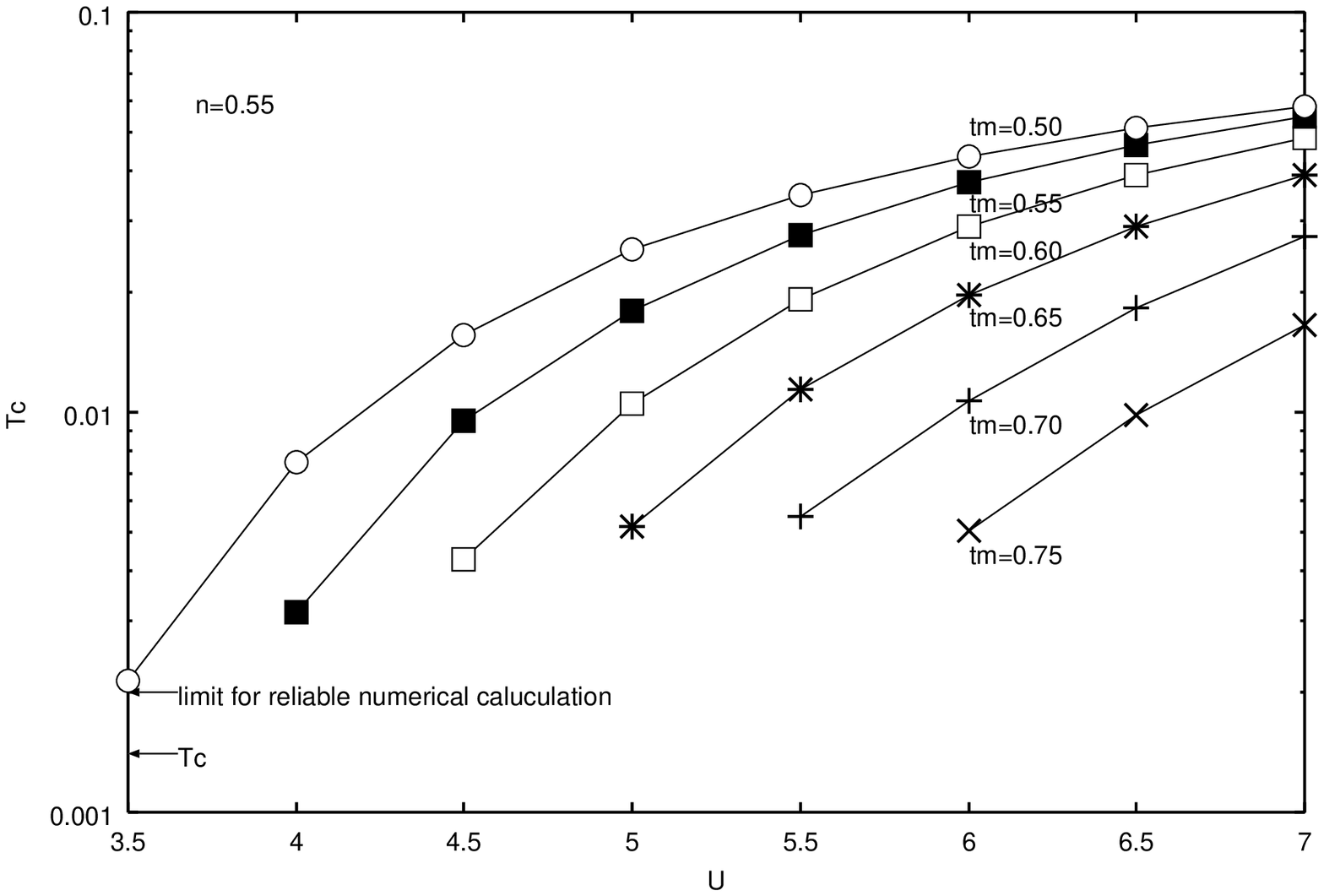}
\caption{The calculated $T_{\rm c}$ as $U$ is varied, at $n=0.55$ and for various 
 values of $t_{\rm m}$ as shown in the figure.}
\label{fig:nTc0.55}
\end{figure}

%8
\begin{figure}
\includegraphics[width=1.0 \linewidth]{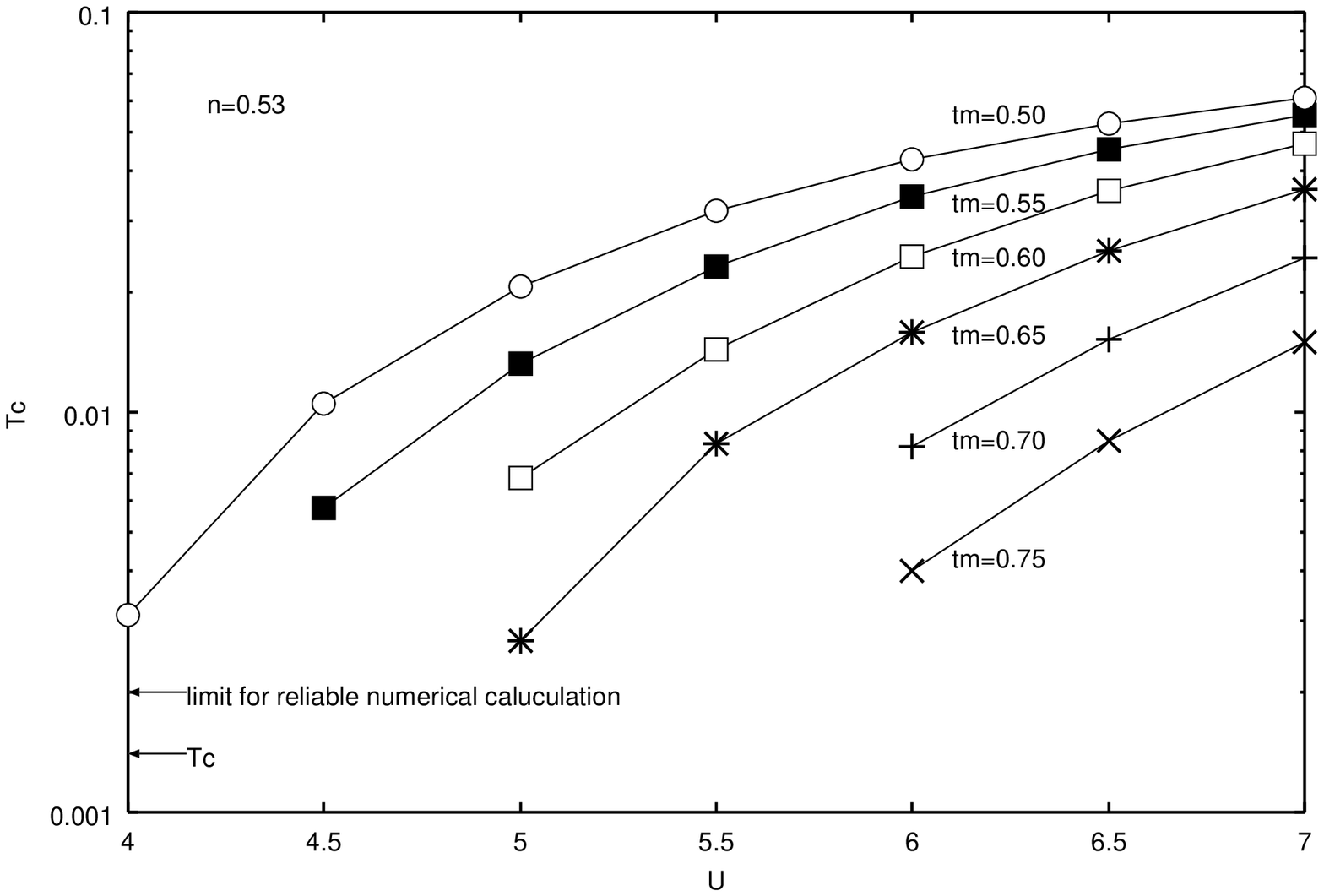}
\caption{The calculated $T_{\rm c}$ as $U$ is varied, at $n=0.53$ and for various 
 values of $t_{\rm m}$ as shown in the figure.}
\label{fig:nTc0.53}
\end{figure}

%9

%10

%11

%12
\begin{figure}
\includegraphics[width=1.0 \linewidth]{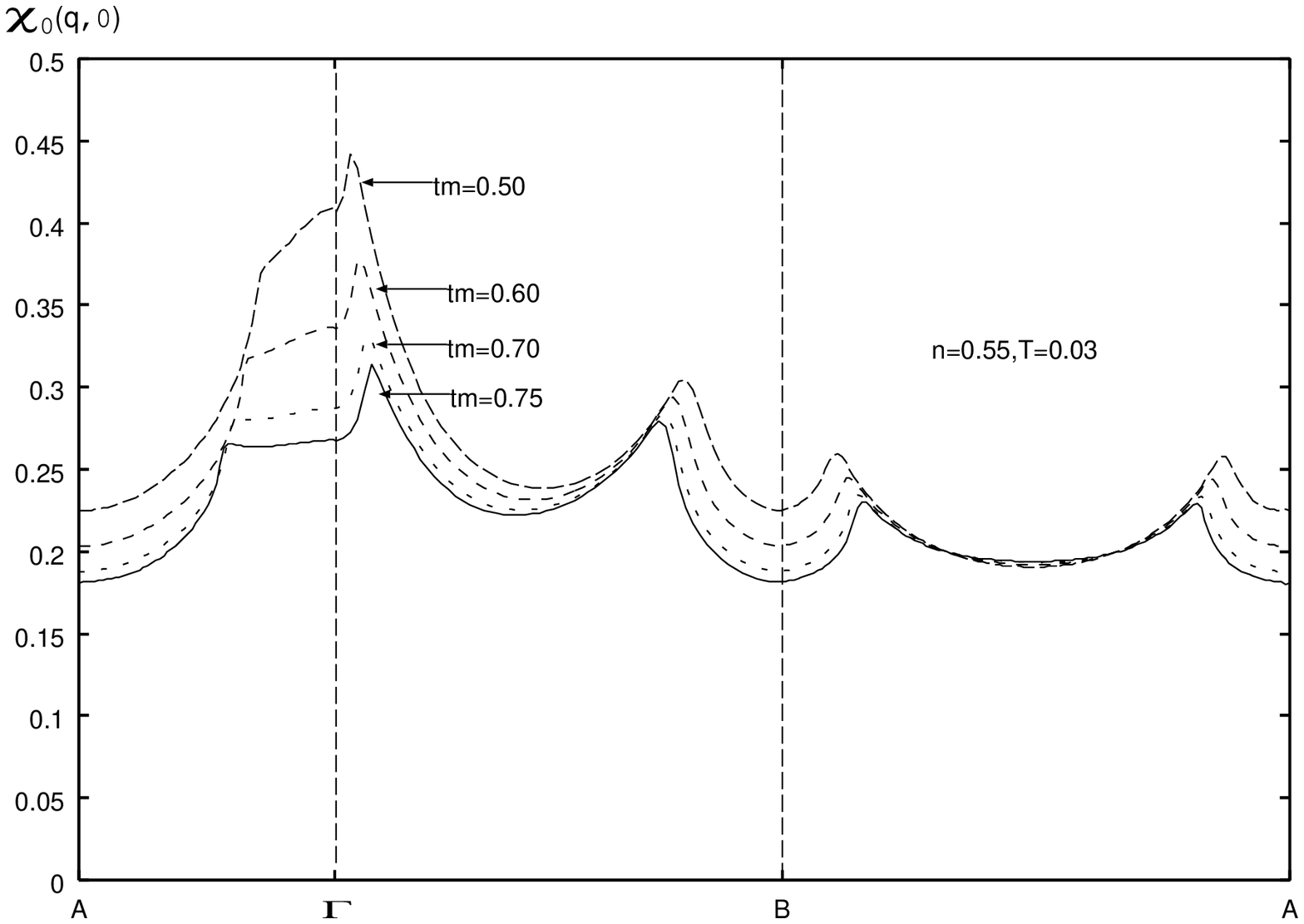}
\caption{The momentum dependence of the static bare susceptibility for
 various $t_{\rm m}$. The notations under the horizontal axis are defined in
 Fig.2. These results are obtained for $n=0.55$, $T=0.03$.}
\label{fig:X0tmn0.55t0.03}
\end{figure}

%13

%14

%15

%16

%17

\end{document}